\documentclass[conference, A4paper]{IEEEtran}
\usepackage{graphicx, amsmath, amsfonts, amssymb, xcolor, subfig, enumerate, mathrsfs,  tikz, standalone, mathtools, bbm}
\usetikzlibrary{arrows, patterns}
  \usepackage{pgfplots}
  \pgfplotsset{compat=newest}
  \usetikzlibrary{plotmarks}
  \usetikzlibrary{arrows.meta}
  \usepgfplotslibrary{patchplots}

\usepackage{amsthm}

\newtheorem{theorem}{Theorem}

\newtheorem{corollary}{Corollary}

\newtheorem{definition}{Definition}

\theoremstyle{definition}

\theoremstyle{remark}
\newtheorem{remark}{Remark}

\global\long\def\EE{\mathbb{E}}
\global\long\def\PP{\mathbb{P}}

\global\long\def\11{\mathbbm{1}}

\global\long\def\+{\oplus}

\newcommand{\sign}{\mathsf{sign}}
\newcommand{\OR}{\mathsf{OR}}

\newcommand{\MAJ}{\mathsf{MAJ}}

\usepackage{stackengine}
\def\deq{\mathrel{\ensurestackMath{\stackon[1pt]{=}{\scriptstyle\Delta}}}}

\newtheorem{lem}{Lemma}

\newcommand{\be}{\begin{equation}}
\newcommand{\ee}{\end{equation}}
\newcommand{\ben}{\begin{equation*}}
\newcommand{\een}{\end{equation*}}
\newcommand{\mc}{\mathcal}

\newcommand{\abs}[1]{\lvert#1\rvert}

\newcommand{\expec}{\mathbb{E}}

\DeclareMathOperator*{\argmax}{arg\,max}
\newcommand{\reals}{\mathbb{R}}
\newcommand{\prob}{\mathbb{P}}

\newcommand{\barf}{\bar{f}}
\def\<{\langle}
\def\>{\rangle}
\newcommand{\bfx}{\boldsymbol{x}}
\newcommand{\bfX}{\boldsymbol{X}}
\newcommand{\bfy}{\boldsymbol{y}}
\newcommand{\bfY}{\boldsymbol{Y}}
\newcommand{\Pek}{\optfont{P^k}}
\newcommand{\Plin}{\optfont{P^{\sf{lin}}}}

\newcommand{\optfont}[1]{\mathsf{#1}}
\newcommand{\fsetfont}[1]{\mathcal{#1}}

\IEEEoverridecommandlockouts

\newcommand\pmm{\{-1,1\}}

\begin{document}

\title{Boolean Functions with Biased Inputs: Approximation and Noise Sensitivity}
\author{
\IEEEauthorblockN{Mohsen Heidari}
\IEEEauthorblockA{University of Michigan, USA \\
\tt mohsenhd@umich.edu
} 
\and
\IEEEauthorblockN{S. Sandeep Pradhan}
\IEEEauthorblockA{University of Michigan, USA \\
\tt pradhanv@umich.edu
}
\and
\IEEEauthorblockN{Ramji Venkataramanan}
\IEEEauthorblockA{University of Cambridge, UK \\
\tt ramji.v@eng.cam.ac.uk  }
\thanks{This work was supported in part by a grant from the Michigan Cambridge Research Initiative (MCRI) and by NSF grant CCF 1717299.}
}
\maketitle

\begin{abstract}
This paper considers the problem of approximating a  Boolean function $f$ using another Boolean function from a specified  class. Two classes of approximating functions are considered: $k$-juntas, and linear Boolean functions. The $n$ input bits of the function are assumed to be independently drawn from  a distribution that may be biased. The quality of approximation is measured by the mismatch probability between $f$ and the approximating function $g$.   For each class, the optimal approximation and the associated mismatch probability is characterized in terms of the biased Fourier expansion of  $f$. The technique used to analyze the mismatch probability also  yields an expression for the noise sensitivity of $f$ in terms of the biased Fourier coefficients, under a general i.i.d. input perturbation model.
\end{abstract}

\section{Introduction} \label{sec:intro}

\IEEEPARstart{G}iven a set of labeled data, we may wish to learn the optimal classifier within a specific class of functions. For example,  given $n$-dimensional data with binary labels, one may wish to construct a classifier that depends on only  $k$ of the $n$ input variables (where $k$ may be much smaller than $n$). Such a parsimonious classifier would be less accurate on the training data than the optimal  unconstrained classifier (which uses all $n$ variables), but may be more robust to errors in the data.  A useful measure to quantify this trade-off is the probability of mismatch between the optimal unconstrained and constrained classifiers, under some distribution on the input variables.

Motivated by such applications, we consider the problem of approximating a given Boolean function $f: \pmm^n \to \pmm$ using a simpler Boolean function from a specified class.  The input set $\pmm^n$ is equipped with a product  distribution, where each of the $n$ input bits $X_1, \ldots, X_n$ is drawn independently according to 
\be \prob(X_i=-1)=1- \prob(X_i=1)=p, \qquad i  \in [n]. \label{eq:Xidist} \ee
The quality of approximation is measured by the \emph{mismatch probability} $\prob(f(\bfX) \neq g(\bfX))$, where $\bfX \deq (X_1, \ldots, X_n)$.

We consider two classes of approximating functions: i) $k$-\emph{juntas} where the Boolean  function $g$ depends on at most $k$ of the $n$ input variables (with $k <n$), and ii) linear Boolean functions which are  parity functions or negations of a parity on a subset of the input variables. In each case, we characterize the optimal approximation and the associated mismatch probability in terms of the $p$-biased Fourier expansion of the original function $f$.

The standard Fourier expansion \cite{o2014analysis} of a Boolean function is a multilinear polynomial with real coefficients, where each term in the polynomial corresponds to  a parity function on a subset of the input variables.  
The Fourier expansion has been  used to analyze Boolean functions  in wide range of applications, e.g., to characterize the learning complexity \cite{linial1993constant,mossel2004learning}, noise sensitivity \cite{o2014analysis,kalai2005noise,LiMedardISIT18}, approximation \cite{blais2010polynomial}, and other information-theoretic properties \cite{courtade2014boolean,weinberger2017optimal,weinberger2018self}.   The parity functions form a set of orthonormal basis functions when the inputs to the Boolean function are uniformly random. 

For $p \in (0,1)$, the $p$-biased Fourier expansion \cite[Chap. 8]{o2014analysis} generalizes the standard Fourier expansion by expressing the Boolean function  as a linear combination of functions that form an orthonormal basis when the input variables are drawn i.i.d. according to the distribution in \eqref{eq:Xidist}.  $p$-biased Fourier analysis was used in \cite{furst1991improved}  to show that  a certain class of Boolean functions could be learnt  efficiently using examples drawn  from a biased input distribution. It has also been used to study threshold phenomena of random graphs \cite{talagrand1994russo}. In this paper, we use the $p$-biased expansion to study optimal approximation of Boolean functions with biased inputs. 

The contributions of the paper are as follows. 
\begin{enumerate}
\item In Section  \ref{sec:bool_joint}, we obtain an expression (Lemma \ref{lem:genXY}) for the mismatch probability 
$\prob(f(\bfX) \neq g(\bfY))$, where  $f,g$ are Boolean functions with statistically dependent binary inputs $\bfX$ and $\bfY$, respectively. Taking $f=g$ yields the noise sensitivity of a Boolean function under a general i.i.d. input perturbation model. Lemma \ref{lem:genXY}
also generalizes  a bound on the mismatch probability obtained in  \cite{farhad2017correlation}.

\item Next, by taking $\bfX= \bfY$, Lemma \ref{lem:genXY}  is  used to establish  the optimal approximation with $k$-juntas (Section \ref{sec:kjunta}),  and with linear Boolean functions (Section \ref{sec:linear}).
We provide examples to illustrate how the optimal approximation within a class depends on the input bias. 
\end{enumerate}
We remark that some of the results (such as those in Section \ref{sec:kjunta}) hold for product distributions over any finite input alphabet. For concreteness, we focus on the binary input alphabet throughout the paper. We also mention that the worst-case circuit-size complexity of approximating Boolean functions with uniform inputs was analyzed in \cite{andreev1997optimal}.

\emph{Notation}: We use $[n]$ to denote the set $\{1, \ldots, n \}$. The cardinality of a set $S$ is denoted by $\abs{S}$. Given  $S\subseteq [n]$ and a sequence of numbers $a_i, i\in [n]$, denote $a^S \deq (a_i)_{i\in S}$. We use upper case to denote random variables, lower case for  realizations, and boldface for vectors. 

\section{The $p$-biased Fourier Expansion}
We consider Boolean functions with the distribution on the entries of the input $\bfX=(X_1, \ldots, X_n)$  being i.i.d. according to  \eqref{eq:Xidist}.
With this distribution, an inner product can be defined  for the (larger) space of bounded functions with binary inputs  and real-valued outputs.  For any  $f, g: \{ -1, 1 \}^n \to \reals$, let
\be
\< f, g \> \deq \expec[f(\bfX)g(\bfX)] = \sum_{\bfx \in \{-1,1 \}^n}  \hspace{-4pt} \prob(\bfX = \bfx) f(\bfx) g(\bfx).
\ee

 The  $p$-biased Fourier expansion \cite[Chap. 8]{o2014analysis} of a function  $f: \{-1, 1\}^n \to \reals$ is 
 \be
f(\bfx) = \sum_{S \subseteq [n]} \barf(S) \phi_S(\bfx),
\label{eq:pbiased_def}
\ee
where \be \phi_S(\bfx) = \prod_{i \in S} \frac{(x_i - \mu)}{\sigma}.  \label{eq:phiS_def} \ee
Here
\be  
\mu = (1-2p) \ \text{ and }  \ \sigma= 2 \sqrt{p(1-p)}
\label{eq:musig_def}
 \ee
are the mean and standard deviation, respectively, of each of the $X_i$'s. For $S \subseteq [n]$, the $p$-biased Fourier coefficients can be computed as 
\be
\barf(S) = \< f, \phi_S \> = \expec[f(\bfX) \phi_S(\bfX)],
\ee
where the entries of $\bfX=(X_1, \ldots, X_n)$ are i.i.d. according to \eqref{eq:Xidist}. Under this inner product, the set of functions $\{ \phi_S \}_{S \subseteq [n]}$ is an orthonormal basis. Indeed, using the  independence of the $X_i$'s, it can be shown that for any $S, T \subseteq [n]$, the inner product $\expec[\phi_S(\bfX) \phi_T(\bfX)] =1$ if $S=T$, and $0$ otherwise.

Since \eqref{eq:pbiased_def} is an orthonormal expansion, the inner product between two functions can be expressed in terms of their $p$-biased Fourier coefficients. For  any $f,g: \{-1, 1\}^n \to \reals$
\be
\< f, g \> = \expec[f(\bfX) g(\bfX)]=  \sum_{S \subseteq [n]} \barf(S) \bar{g}(S).
\ee

The standard Fourier expansion corresponds to the case where $p=\frac{1}{2}$. In this case, $\mu=0, \sigma=1$, and the basis functions are  $\phi_S(\bfx) = \prod_{i \in S} x_i$, $S \subseteq [n]$.

For $f: \{-1, 1\}^n \to \reals$ and any set $T \subseteq [n]$, let $\bfX^T$ denote the components of $\bfX$ indexed by $T$. We refer to $\EE[f | \bfX^T]$ as the projection of $f$ onto $\bfX^T$. This projection 
 is denoted by $f^{\subseteq T}$, and is given by 
\be
\label{eq:fprojT}
f^{\subseteq T}(\bfX) \deq \EE[f(\bfX) | \bfX^T] = \sum_{S \subseteq T} \barf(S) \phi_S(\bfX^S).
\ee
The last equality is obtained  from  \eqref{eq:pbiased_def} by noting that for any set $S \not\subseteq T$, the conditional expectation $\expec[\phi_S(\bfX)  \, |  \, \bfX^T]=0$. We note that the projection $f^{\subseteq T} $ may have real-valued outputs, even when $f$ is Boolean.


\section{Boolean functions of jointly distributed  random variables} \label{sec:bool_joint}

In this section we investigate Boolean functions, say $f$ and $g$, whose inputs that are statistically correlated. We derive an expression for the mismatch probability in term of biased Fourier coefficients of the functions.

Let $X, Y \in \{ -1,1 \}$ be jointly distributed Boolean random variables with  joint pmf $P_{XY}$ whose marginals satisfy
\begin{align}
\prob(X=-1)=p, \quad \prob(Y=-1)=q.
\end{align}
Let $\rho \in [-1,1]$ denote the correlation coefficient between $X$ and $Y$. The joint pmf $P_{XY}$ is uniquely determined by the triple $(p,q,\rho)$. Let 
$(\bfX, \bfY)$ be a pair of sequences with  entries $(X_i, Y_i)_{i \in [n]} \, \sim_{i.i.d.} \, P_{XY}$.

For any Boolean functions  $f, g:\{-1,1 \}^n \to \{-1, 1 \}$,  the $p$-biased Fourier expansion of $f$  is given by \eqref{eq:pbiased_def}--\eqref{eq:phiS_def}, and the $q$-biased Fourier expansion of  $g$ is
\begin{align}
g(\bfy) = \sum_{S \subseteq [n]} \tilde{g}(S) \psi_S(\bfy),
\end{align}
where
$\psi_S(\bfy) = \prod_{i \in S} \frac{ (y_i - \mu')}{\sigma'},$
with 
\be
\mu'= (1-2q), \quad \sigma'=2\sqrt{q(1-q)}.
\ee
The $q$-biased Fourier coefficients of $g$ are
\be
\tilde{g}(S) = \<g, \psi_S \> = \expec[ g(\bfY) \psi_S(\bfY)],  \qquad \forall S \subseteq [n].
\ee

The following result expresses the probability of mismatch between $f(\bfX)$ and $g(\bfY)$ in terms of their biased Fourier coefficients.
\begin{lem}\label{lem:genXY}
For $(\bfX, \bfY)$ with $(X_i, Y_i)_{i \in [n]} \, \sim_{i.i.d.} \, P_{XY}$, 
\begin{align}
\expec[ f(\bfX) g(\bfY) ] & =\sum_{S \subseteq [n]} \barf(S)  \tilde{g}(S) \rho^{\abs{S}}, \label{eq:EfXgY} \\
\prob(f(\bfX ) \neq g(\bfY)) &  = \frac{1}{2} - \frac{1}{2} \sum_{S \subseteq [n]} \barf(S)  \tilde{g}(S) \rho^{\abs{S}}. \label{eq:prob:fxgy_mismatch}
\end{align}
\end{lem}
\begin{IEEEproof}
Using the $p$-biased Fourier expansion for $f(\bfX)$ and the $q$-biased one for $g(\bfY)$, we have
\begin{align}
& \expec[ f(\bfX) g(\bfY) ]  = \sum_{S \subseteq [n], S' \subseteq [n] } \barf(S)  \tilde{g}(S')   \expec[ \phi_S(\bfX) \psi_{S'}(\bfY) ] \nonumber \\
& =  \sum_{S \subseteq [n], S' \subseteq [n] } \barf(S)  \tilde{g}(S') \expec \Big[ \prod_{i \in S, j \in S'}  
\Big(\frac{X_i - \mu}{\sigma} \Big) 
\Big( \frac{Y_j - \mu'}{\sigma'} \Big) \Big] \nonumber \\
& \stackrel{(a)}{=}  \sum_{S \subseteq [n]}  \barf(S)  \tilde{g}(S)  \prod_{i \in S} \expec \Big[ \Big(\frac{X_i - \mu}{\sigma} \Big) 
\Big( \frac{Y_i - \mu'}{\sigma'} \Big)  \Big] \nonumber \\
& =  \sum_{S \subseteq [n]}  \barf(S)  \tilde{g}(S) \rho^{\abs{S}}.
\end{align}
Here $(a)$ is obtained as follows, using the independence of  the $(X_i, Y_i)$ pairs across $i \in [n]$: when $S \neq S'$, there is at least one index that belongs to only one of these two sets. If $i \in S$ and $i \notin S'$, the term $\expec[(X_i - \mu)/\sigma] =0$; similarly if $j \in S'$ and $j\notin S$, then  $\expec[(Y_j - \mu')/\sigma'] =0$.

Eq. \eqref{eq:prob:fxgy_mismatch} follows by observing that
\[  \expec[ f(\bfX) g(\bfY) ] = 1 \cdot \prob(\bfX = \bfY)  - 1 \cdot \prob(\bfX \neq \bfY). \]
\end{IEEEproof}

For $\abs{\rho} <1$, Lemma \ref{lem:genXY} shows that the biased Fourier coefficients corresponding to sets of small cardinality play a key role in determining probability of mismatch.  Since $f$ and $g$ are Boolean, by  Parseval's formula we have
\be
\sum_{S \in [n]} \abs{\bar{f}(S)}^2 = \sum_{S \in [n]} \abs{\tilde{g}(S)}^2= 1.
\ee
Suppose that the biased Fourier coefficients of $f$ and $g$ are both largely concentrated on sets $S$ of small cardinality. Then if the coefficients have the same sign on these sets, then \eqref{eq:prob:fxgy_mismatch} shows that the probability of mismatch between $f(\bfX)$ and $g(\bfY)$ will be small; if the coefficients have opposite signs on these sets, the probability of mismatch will be close to $1$.  On the other hand, if the biased Fourier coefficients of 
$f, g$ are concentrated on sets $S$ of large cardinality, then for $\rho<1$, the probability of mismatch will be close to $1/2$.

\emph{Noise sensitivity}: The noise sensitivity of a Boolean function $f: \pmm^n \to \pmm$  is defined as  $\prob(f(\bfX) \neq f(\bfY))$, where $(X_i, Y_i)_{i \in [n]} \sim_{i.i.d.} P_{XY}$. It represents the mismatch probability under a perturbation model where the noisy  input $\bfY$ is assumed to be generated from the original  input $\bfX \sim_{i.i.d.}P_X$ via a memoryless channel $P_{Y|X}$. 

 By taking $f=g$, Lemma \ref{lem:genXY} yields the noise sensitivity for a general bivariate   distribution $P_{XY}$ on a pair of Boolean random variables, parametrized by $(p, q, \rho)$. From \eqref{eq:prob:fxgy_mismatch}, the noise sensitivity of $f$ can be expressed as
 \be
 \optfont{NS}_{(p,q,\rho)} = 
  \frac{1}{2} - \frac{1}{2} \sum_{S \in [n]} \barf(S)  \tilde{f}(S) \rho^{\abs{S}},
 \ee
where $\barf(S)$ and $ \tilde{f}(S)$ are the $p$-biased and $q$-biased Fourier coefficients, respectively. This generalizes previous characterizations of noise sensitivity \cite{o2014analysis,blais2010polynomial}, which assumed a symmetric perturbation model with $p=q$.



In the following sections, we will use Lemma \ref{lem:genXY}  to  obtain the mismatch probability for approximations of Boolean functions. We will apply Lemma \ref{lem:genXY}   taking $g$ to be the approximating function, and with $\bfX= \bfY$ (i.e., $\rho=1$).

\section{Approximation with $k$-Juntas} \label{sec:kjunta}

In the set of Boolean functions with $n$ input variables, $k$-juntas are Boolean functions whose output depends only on a subset of at most $k$ input variables.  
\begin{definition} 
A Boolean function $g: \{-1,1 \}^n \to \{-1, 1\}$  is a $k$-junta (with $k< n$), if there exist $i_1, i_2, ..., i_k\in [n]$ and a Boolean function $h: \pmm^k\mapsto \pmm$ such that 
\begin{equation*}
g(\bfx)=h(x_{i_1}, x_{i_2}, ..., x_{i_k}), \quad \forall \bfx\in \pmm^n.
\end{equation*} 
\end{definition}

In this section, we investigate approximation of Boolean functions by $k$-juntas.
 Given a Boolean function $f: \pmm^n \mapsto \{-1,1\}$, we wish to find a $k$-junta $g$ that  minimizes the mismatch probability 
$\PP(f(\bfX)\neq g(\bfX)),$
where the entries of $\bfX=(X_1, \ldots, X_n)$ are i.i.d. according to \eqref{eq:Xidist}. Letting $\fsetfont{B}_k$ denote the set of all $k$-juntas, the minimum mismatch probability is denoted by
\begin{equation}
\Pek[f] \deq \min_{g\in \fsetfont{B}_k} \PP (f(\bfX)\neq g(\bfX)).
\end{equation}

The following theorem gives an expression for $\Pek[f]$ and an optimal $k$-junta function for approximation of $f$. For $x \in \reals$, we define $\sign(x)=1$ if $x \geq 0$, and $-1$ if $x <0$.

\begin{theorem} \label{thm:kjunta}
Let  $f:\pmm^n  \to \pmm$ be a Boolean function with input $ \bfX= (X_i)_{i \in [n]}$ i.i.d.
according to the distribution in \eqref{eq:pbiased_def}. Then the minimum mismatch probability of a $k$-junta approximation of $f$ (for $k< n$) is 
\begin{equation}\label{eq:minimum_mismatch_prob}
 \Pek[f] = \frac{1}{2}\left[  1-\max_{J\subseteq [n], \,  \abs{J} \leq k}\| \,  f^{\subseteq J}\|_1 \right],
\end{equation}
where $f^{\subseteq J}$ is the projection defined in \eqref{eq:fprojT}, and 
\be
\|f^{\subseteq J}\|_1 =  \EE[ \, \abs{f^{\subseteq J}(\bfX)} \, ].  
\ee
Furthermore, the minimum mismatch probability is achieved by the $k$-junta approximation 
$g= \sign(f^{\subseteq J^*})$,
 where $J^*$ achieves the optimum in \eqref{eq:minimum_mismatch_prob}. 
\end{theorem} 
\begin{IEEEproof}
We apply Lemma  \ref{lem:genXY} taking $g$ to be a $k$-junta, and $\rho=1$, i.e., $\bfX=\bfY$. From \eqref{eq:prob:fxgy_mismatch},  for any $g$  the mismatch probability satisfies
\begin{equation}\label{eq:k-junta:mismatch}
\prob(f(\bfX ) \neq g(\bfX))   = \frac{1}{2} - \frac{1}{2} \sum_{S \in [n]} \barf(S)  \bar{g}(S),
\end{equation}
where $\barf(S), \bar{g}(S)$ are the $p$-biased Fourier coefficients of $f$ and $g$, respectively.
Suppose that $g(\bfx)$ depends on the inputs $(x_i)_{i\in J}$, where $J$ is a subset of $[n]$ with at most $k$ elements. Then, $\bar{g}(S)=0$ for any $S \not \subseteq J$. Hence, the mismatch probability in \eqref{eq:k-junta:mismatch} equals 
\begin{align}
\nonumber
& \prob(f(\bfX ) \neq g(\bfX))   = \frac{1}{2} - \frac{1}{2} \sum_{S \subseteq J} \barf(S)  \bar{g}(S) = \frac{1}{2} - \frac{1}{2}\<f^{\subseteq J},g \>\\ 
&   \geq \frac{1}{2} - \frac{1}{2}
\<  \abs{f^{\subseteq J}}, \abs{g} \> =   \frac{1}{2} - \frac{1}{2} \|f ^{\subseteq J} \|_1.
\label{eq:k-junta:mismatch 2}
\end{align}
The last equality in \eqref{eq:k-junta:mismatch 2} holds because $g$ is a Boolean function, hence $\|g\|=1$. Since $J$ is an arbitrary subset of $[n]$ with at most $k$ elements, \eqref{eq:k-junta:mismatch 2} implies
\begin{align}
\label{eq:Pek_lb}
 \Pek[f] \geq \frac{1}{2}-\frac{1}{2} \max_{J\subseteq [n], \,  \abs{J} \leq k} \|f^{\subseteq J}\|_1.
\end{align}

Next we obtain an upper bound on  $\Pek[f]$ by specifying a $k$-junta approximation of $f$.  Fix a subset  $J\subseteq [n]$ with $|J|\leq k$, and let $g=\sign[f^{\subseteq J}]$. Note that for any $f$ we have 
\begin{align} 
\nonumber
\<g, f^{\subseteq J}\> & = \expec[ f^{\subseteq J}(\bfx) \cdot \sign(f^{\subseteq J}(\bfx)) ] \\
& = \EE[ \, \abs{f^{\subseteq J}(\bfx)} \, ]  =\|f^{\subseteq J}\|_1. \label{eq:innerL1}
\end{align} 
Therefore, using \eqref{eq:k-junta:mismatch 2}, the mismatch probability of this approximation is
\begin{equation}
\prob(f(\bfX ) \neq g(\bfX))=  \frac{1}{2}-\frac{1}{2}\|f^{\subseteq J}\|_1.
\label{eq:mis_achieve}
\end{equation}
Eq. \eqref{eq:mis_achieve}  provides an upper-bound on $ \Pek[f]$ for any $J$ such that $\abs{J} \leq k$. Taking $J=J^*$, where $J^*$ achieves $ \max_{J\subseteq [n], \,  \abs{J} \leq k}\|f^{\subseteq J}\|_1$, we obtain
 \begin{equation}
 \Pek[f]  \leq  \frac{1}{2}-\frac{1}{2} \max_{J\subseteq [n], \,  \abs{J} \leq k} 
 \|f^{\subseteq J}\|_1.
 \label{eq:pekub}
 \end{equation}
 Combining \eqref{eq:pekub} and \eqref{eq:Pek_lb} completes the proof.
 \end{IEEEproof}

\begin{remark}
The proof shows that for any $J \subseteq [n]$, the mismatch probability between $f$ and  $\sign[f^{\subseteq J}]$ is given by \eqref{eq:mis_achieve}. The function $\sign[f^{\subseteq J}]$  is the maximum a posteriori probability (MAP) estimator 
 of $f$ given $J$.  To see this, note that the MAP estimator of $f$ given $\bfX^J$ is a Boolean function $g$ such that $g(\bfx^J)=1$ if 
$$\prob( f(\bfX)=1| \bfX^J=x^J)  \geq \prob(f(\bfX)=-1| \bfX^J=\bfx^J),$$ and $g(\bfx^J)=-1$ otherwise.
Since $f$ is a Boolean function, by the definition of $f^{\subseteq J}$, we have $$f^{\subseteq J}(\bfX^J) =\EE[f| \bfX^J]=\prob\{f=1| \bfX^J\}-\prob\{f=-1| \bfX^J\}.$$
Hence, $\sign[f^{\subseteq J}]$ equals the MAP estimator of $f$.
\end{remark}


\begin{remark}
Eq. \eqref{eq:mis_achieve}  shows that the mismatch probability for approximating $f$ with $\sign[f^{\subseteq J}]$ is determined by $\| f^{\subseteq J} \|_1$. We can bound the mismatch probability from above and below in terms of $\| f^{\subseteq J} \|_2$, which depends only the weight of the $p$-biased Fourier coefficients of  $S \subseteq J$.
\end{remark}

\begin{corollary}\label{corr:bounds on mismatch}
With the assumptions of Theorem \ref{thm:kjunta}, the minimum mismatch probability satisfies
\begin{align}
& \frac{1}{2}\left[  1-\max_{J\subseteq [1,n], \,  \abs{J} \leq k}\|f^{\subseteq J}\|_2 \right] 
\nonumber  \\
 &\quad \leq  \Pek[f] \leq  \frac{1}{2}\left[ 1-\max_{J\subseteq [1,n],~ |J|\leq k}\|f^{\subseteq J}\|^2_2\right],
\end{align}
where
\be
\|f^{\subseteq J}\|_2^2 = \EE[f^{\subseteq J}(\bfX)^2] = \sum_{S \subseteq J} \abs{\barf(S)}^2.
\ee
\end{corollary}
\begin{IEEEproof}
Since $-1\leq f^{\subseteq J}\leq 1$, we have $|f^{\subseteq J}(\bfx)|\geq |f^{\subseteq J}(\bfx)|^2 $. Thus $\|f^{\subseteq J}\|_1\geq \|f^{\subseteq J}\|^2_2$, which yields the upper bound by substituting in \eqref{eq:minimum_mismatch_prob}.
Next, from Jensen's inequality we have
\begin{equation*}
\EE\Big[|f^{\subseteq J}(\bfX)|^2\Big] \geq  \Big[ \EE \abs{f^{\subseteq J}(\bfX)}\Big]^2.
\end{equation*}
This implies that $\|f^{\subseteq J}\|_2 \geq \|f^{\subseteq J}\|_1$, which establishes the lower-bound.
\end{IEEEproof}

Given $k <n$,  Theorem \ref{thm:kjunta} specifies the optimal $k$-junta approximation for $f$. The problem may be viewed from another perspective: given  $\epsilon>0$,  find the smallest $k$ such that there exists a $k$-junta function whose mismatch probability with $f$ is at most $\epsilon$. When $f$ depends on all $n$ input variables, there is a trade-off between $k$ and $\epsilon$: the lower the tolerance $\epsilon$, the larger the required value of $k$. As discussed in Section \ref{sec:conc}, this  formulation can be useful in the context of {learning} arbitrary Boolean functions to within a specified mismatch probability. 

\begin{figure}[t]
\centering
\includegraphics[width=3.45in]{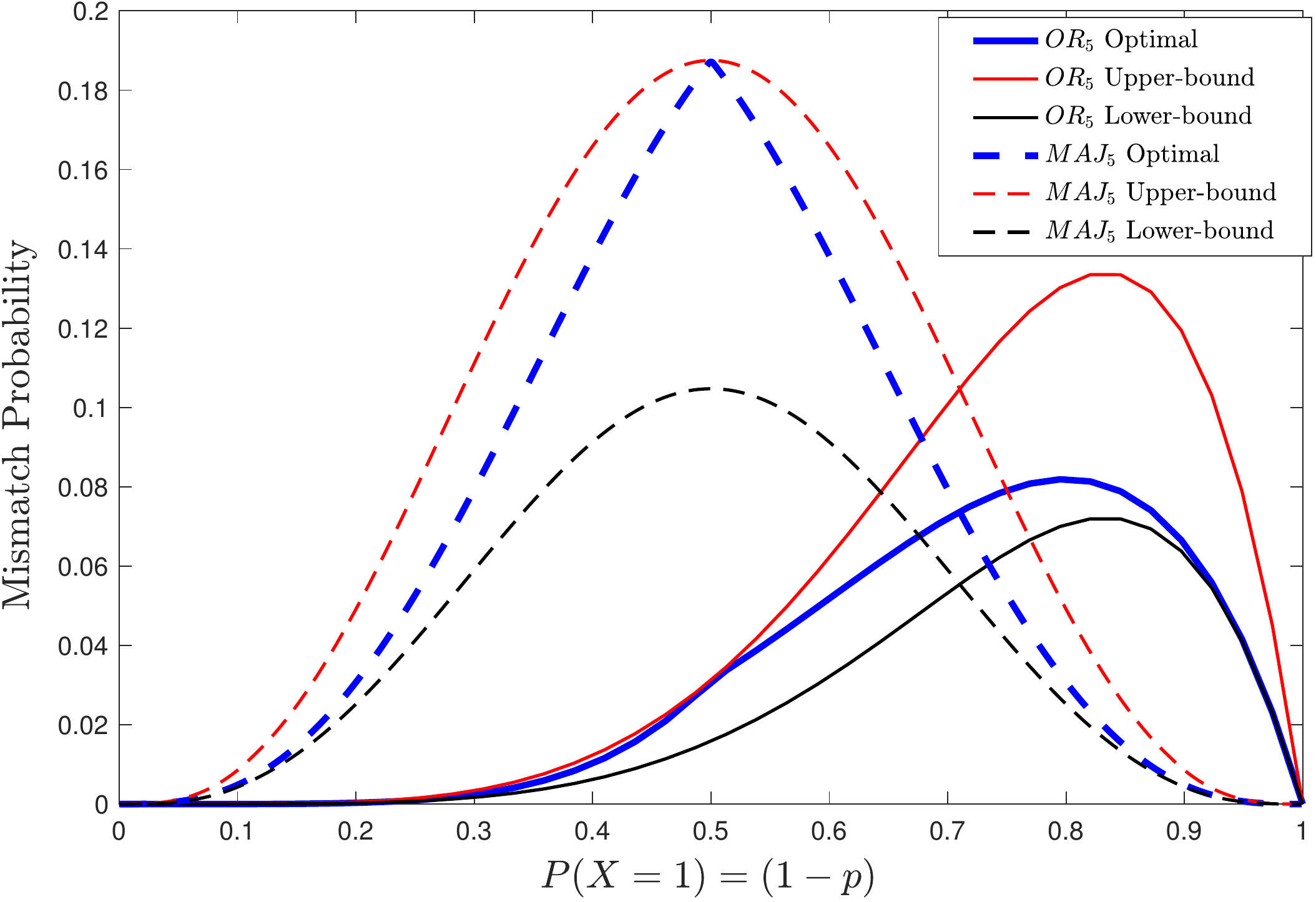}
\caption{Minimum mismatch probability for approximation of $\OR_5$ and $\MAJ_5$ using $4$-juntas.}
\label{fig:OR5 appx}
\vspace{-7pt}
\end{figure}

\emph{Examples}: We examine $k$-junta approximations of  the `or' function $\OR_n$, and the majority function $\MAJ_n$. The function $\OR_n: \{-1,1\}^n \mapsto \{-1,1\}$   is defined as $\OR_n(\bfx)=1$ if $\bfx=(1,1,\ldots, 1)$, and  $\OR_n(\bfx)=-1$ otherwise. The  majority function  is defined as  $\MAJ_n(\bfx)=\sign(\sum_{i=1}^n x_i)$ for all $\bfx\in \pmm^n$.  Figure \ref{fig:OR5 appx} shows the minimum mismatch probability as function of $P_X(1)=(1-p)$ for the approximation of $\OR_5$ and $\MAJ_5$ using $4$-juntas (i.e., $n=5$ and $k=4$). The bounds given in Corollary \ref{corr:bounds on mismatch} are also plotted.  

Using the symmetry between the inputs, we can show that
$$\OR_n^{\subseteq J}(\bfx ^J)=-1+ (1-p)^{n-|J|}\left[1+\OR_{\abs{J}}(\bfx^J)\right].$$
For $(1-p) <\frac{1}{2}$, the optimal approximation $\sign(\OR_n^{\subseteq J})$   
is therefore  the constant function $-1$ (for $\abs{J} <n$).  For $\MAJ_n$, the projection   does not have a compact closed form expression and is computed as 
\[ 
\MAJ^{\subseteq J}_n = \EE \Big[ \sign\Big(\sum_{i=1}^n X_i \Big) \mid \bfX^J \Big].
\]

\vspace{-2pt}
\section{Approximation with linear Boolean functions} \label{sec:linear}

A linear Boolean function is either a parity or a negation of a parity.
%
%
More precisely, a Boolean function $f: \{-1,+1\}^n \mapsto \{-1,+1\}$ is  linear if it is of the form $f(\bfx)= c \, \prod_{i\in S}x_i$ for some subset $S\subseteq [1,n]$ and constant $c \in \{-1,1\}$.  
%

Given a Boolean function $f$, we wish to find a linear Boolean function $g$ that  minimizes the mismatch probability 
$\PP(f(\bfX)\neq g(\bfX))$.
Let $\fsetfont{L}_n$ denote the set of linear Boolean functions with $n$ input variables. The minimum mismatch probability is denoted by
\begin{equation}
\Plin[f] \deq \min_{g\in \fsetfont{L}_n} \PP(f(\bfX)\neq g(\bfX)),
\end{equation}
where the entries of $\bfX=(X_i)_{i \in [n]}$ are i.i.d. according to \eqref{eq:Xidist}.

For any Boolean function $f$ and $S \subseteq [n]$, let
\be
I_S[f] \deq \sum_{S'\subseteq S} \bar{f}(S')\sigma^{|S'|}\mu^{|S|-|S'|},  \label{eq:IS_def}
\ee
where $\mu, \sigma$ are the mean and standard deviation of the $(X_i)_{i \in [n]}$, defined in \eqref{eq:musig_def}.

\begin{theorem}
\label{thm:bool_mismatch}
Let  $f:\pmm^n  \to \pmm$ be a Boolean function with input $ \bfX= (X_i)_{i \in [n]}$ i.i.d.
according to the distribution in \eqref{eq:pbiased_def}. Then the  linear Boolean function   
$g(\bfx)=c^* \, \bfx^{S^*}$
minimizes the mismatch probability  where
\be 
S^* =\argmax_{S \subseteq [n]} \, \abs{I_S[f]}, \text{ and } \   c^*=\sign(I_{S^*}[f]). 
\label{eq:Sc_star}
\ee
The minimum mismatch probability is
$\Plin[f] =\frac{1- \abs{I_{S^*}[f]}}{2}. $
\end{theorem}
\begin{IEEEproof}
We apply Lemma  \ref{lem:genXY} with $\rho=1$ (i.e., $\bfX =\bfY$), and  $g$ a linear Boolean function. From \eqref{eq:EfXgY}--\eqref{eq:prob:fxgy_mismatch},   we have
\begin{equation}
\label{eq:bool_mismatch}
%
\prob(f(\bfX ) \neq g(\bfX))   = \frac{1}{2} - \frac{1}{2} \EE[f(\bfX ) g(\bfX)].
\end{equation}
Since $g$ is linear Boolean, $g(\bfx) = c \,\bfx^{S}$, for some $S \subseteq [n]$, and $c\in \pmm$. Thus
\begin{align}
& \EE[ f(\bfX ) g(\bfX)]  = \EE[  f(\bfX ) \, c \bfX^{S} ]  
= c  \sum_{S' \subseteq S} \barf(S) \EE[ \phi_{S'}(\bfX)\bfX^{S}] \nonumber \\
& = c  \sum_{S' \subseteq S} \barf(S) \prod_{i \in S'}  \EE \left[  \frac{(X_i - \mu)X_i}{\sigma} \right]  \prod_{i \in S\backslash S'} \EE X_i \nonumber \\
& = c \sum_{S' \subseteq S} \barf(S) \sigma^{\abs{S'}} \mu^{\abs{S \backslash S'}} = c\,  I_S[f].
 \end{align}
 Substituting in \eqref{eq:bool_mismatch}, we deduce 
  \be
 \prob(f(\bfX ) \neq g(\bfX))   = ({1 - \, c \, I_S[f]})/{2}.
 \label{eq:bool_mism2}
 \ee
 The mismatch probability in \eqref{eq:bool_mism2} is minimized by taking $S=S^*$ and $c=c^*$, where
 $ S^* =\argmax_{S \subseteq [n]} \, \abs{I_S[f]}$, and  $c^*=\sign(I_{S^*}[f])$. 
\end{IEEEproof}

For uniformly random inputs ($p=\frac{1}{2}$), we have  $\mu=0, \sigma=1$, which implies $I_S[f] = \barf(S)$. The optimal linear approximation can be succinctly characterized in this case.
\begin{corollary}
If the inputs $(X_i)_{i \in [n]}$ are uniformly random, then   the mismatch probability with $f$ is minimized by the linear Boolean function $g(\bfx) = c^*  \bfx^{S^*}$ with 
\be
S^* =\argmax_{S \subseteq [n]} \, \abs{\barf(S)}, \text{ and } \   c^*=\sign(\barf(S)).
\ee
Here $\barf(S)$ is the standard Fourier coefficient for the set $S$.
\end{corollary}

\begin{figure}[t]
\centering
\includegraphics[width=3.45in]{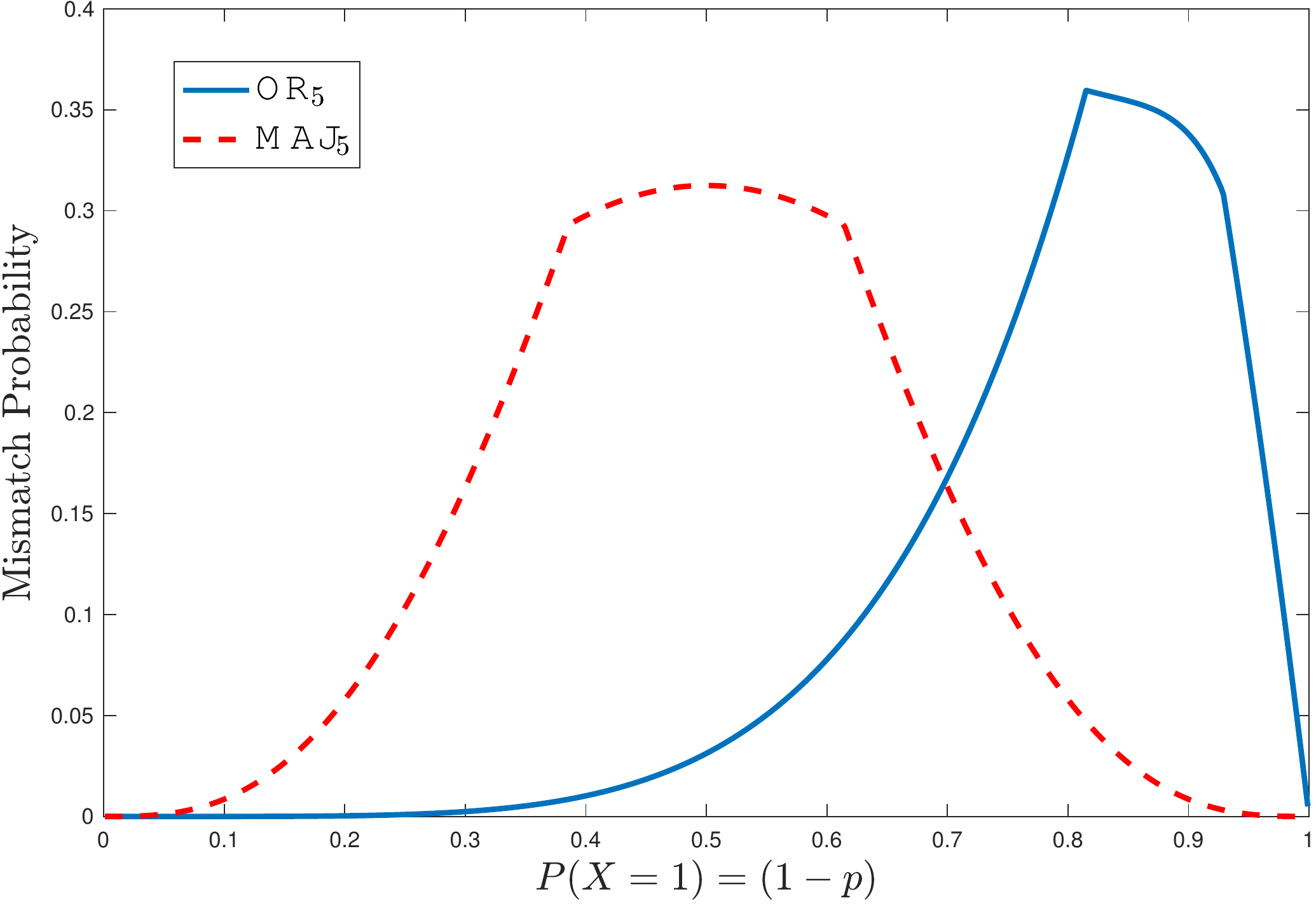}
\caption{Mismatch probability for approximation of $\MAJ_5$ and $\OR_5$ using linear Boolean functions.}
\label{fig:lin_appx}
\vspace{-6pt}
\end{figure}

Figure \ref{fig:lin_appx} shows $\Plin$ for $\OR_5$ and $\MAJ_5$ as a function of $P_X(1)$. The optimal linear approximation for $\OR_5$ is found to be a degree 5 linear function for $P_X(1) \in [0.815, 0.927]$, and the constant  $-1$ function for other values of $P_X(1)$.  For $\MAJ_5$, the optimal linear approximation is a degree $1$ function  for $P_X(1) \in [0.389, 0.611]$,  the constant $-1$ function for $P_X(1) < 0.389$, and the constant $1$ for  
$P_X(1) >0.611$. (The end points of these intervals are accurate up to 3 decimal places.)

\section{Discussion and Future Work} \label{sec:conc}

An interesting open question is whether we can efficiently  \emph{learn} the optimal approximation of an unknown function, using a small  (polynomial in $n$) number of samples from the function. These samples may be generated from either  uniformly distributed or biased inputs. For example, we may wish to learn  the optimal $k$-junta approximation of a function, where $k$ is large enough to achieve a desired mismatch probability. It is known that any $k$-junta  can be learned with high probability with complexity of order $n^{\alpha k+ \mc{O}(1)}$, where $\alpha <1$ \cite{mossel2004learning}. However this result is for the setting where the learning algorithm uses examples from the $k$-junta function. The question of how to efficiently  learn the optimal $k$-junta approximation using examples from the original function is open.  Similar questions may be posed  for other useful classes of approximating functions such as linear threshold functions.


\end{document}